\documentclass[12pt]{iopart}
\usepackage{graphicx,latexsym,stmaryrd,iopams}
 
\begin{document}

\title[1D anyons]{One-dimensional anyons with competing $\delta$-function and 
derivative $\delta$-function potentials}

\author{M. T. Batchelor$^{1,2}$, X.-W. Guan$^1$ and A. Kundu$^3$}
\address{$^1$  
{\small Department of Theoretical Physics, Research School of Physical Sciences and Engineering, 
Australian National University, Canberra ACT 0200, Australia}}
\address{$^2$
{\small Mathematical Sciences Institute, Australian National University, Canberra ACT 0200, Australia}}
\address{$^3$
{\small Saha Institute of Nuclear Physics, Theory Group, 
CAMCS 1/AF Bidhan, Calcutta 700064, India}}

\eads{\mailto{Murray.Batchelor@anu.edu.au}, \mailto{xwe105@rsphysse.anu.edu.au} 
and \mailto{anjan.kundu@saha.ac.in}}

\date{\today}

\begin{abstract}
We propose an exactly solvable model of one-dimensional anyons with competing 
$\delta$-function and derivative $\delta$-function interaction potentials. 
The Bethe ansatz equations are derived in terms of the $N$-particle sector for the
quantum anyonic field model of the generalized derivative nonlinear Schr\"{o}dinger equation.  
This more general anyon model exhibits richer physics than that of the recently studied 
one-dimensional model of $\delta$-function interacting anyons.
We show that the anyonic signature is inextricably related to the velocities of the colliding particles
and the pairwise dynamical interaction between particles.
\end{abstract}

\pacs{04.20.Jb, 05.30.Pr}

\ams{82B21, 82B23}


\section{Introduction}

Anyons -- which interpolate between bosons and fermions -- may exist in
two dimensions (2D), obeying fractional statistics. 
Fractional statistics have recently been observed in experiments on the
quasiparticle excitations of a 2D electron gas in the fractional
quantum Hall (FQH) regime \cite{cond1}. 
The anyonic quasiparticles have profound implications for topological quantum states of matter \cite{QFHE}. 
For example, non-Abelian topological order can be studied through manipulating FQH states in a 2D  
electron gas.
As a consequence, the concept of anyons has become important in
topological quantum computation \cite{comp1,comp2}. 
In one dimension (1D) anyons acquire a step-function-like phase when two
identical particles exchange their positions. 
Different aspects of anyons in 1D have been considered \cite{1D-anyon1,1D-anyon2,1D-anyon3}. 
Among these an anyonic extension \cite{Kundu} of the 1D integrable Bose gas with $\delta$-function
interaction \cite{LL,McG} has attracted recent attention,  including the basic construction of 
the anyon model \cite{BGH, Korepin-1,Zhu}.
Many new results have been obtained for this model, including the connection with 
Haldane \cite{Haldane} exclusion statistics \cite{BGO,BG}, 
correlation functions \cite{Korepin-1,Korepin-2,Satachiara1,Calabrese}, 
entanglement \cite{Satachiara2} and expanding 
anyonic fluids \cite{delCampo}.\footnote{Further 
discussion of the ground state properties is given in Ref. \cite{HZC}.}

A key feature of 1D anyons is that they retain
fractional statistics in quasi-momentum space and the anyonic signature of the exchange interaction  -- 
the topological anyonic interaction and dynamical interaction are
inextricably related.  
1D anyons show an intriguing sensitivity of the boundary condition of their wave  function on the
specific position of the particles \cite{BGH,Korepin-1,AN}.
Therefore the study of exactly solvable interacting
anyons in 1D should give insight into understanding
topological effects in many-body physics.

In this communication we propose a 1D model of anyons with $\delta$-function and 
derivative $\delta$-function interaction and solve the model by means of the co-ordinate Bethe ansatz.
The model contains an additional independent interaction parameter so that in principle one can 
study anyonic signatures through tuning the interactions. 
For this more general model 
we show that the anyonic signature is inextricably related to the velocities of the colliding
particles and the pairwise dynamical interaction between identical particles.

\section{The model}

We consider creation and annihilation operators $\Psi^{\dagger}(x)$ and $\Psi(x)$ at point $x$, 
which satisfy the anyonic commutation relations
\begin{eqnarray}
\Psi (x) \Psi^{\dagger}(y)&=&{\mathrm e}^{-\rmi \kappa w(x,y)}\Psi ^{\dagger}(y)\Psi(x)+\delta(x-y) \nonumber\\
\Psi^{\dagger}(x) \Psi^{\dagger}(y)&=&{\mathrm e}^{\rmi \kappa w(x,y)} \Psi^{\dagger}(y)\Psi^{\dagger}(x) \nonumber\\
\Psi(x) \Psi(y)&=&{\mathrm e}^{\rmi \kappa w(x,y)} \Psi(y)\Psi(x).
\end{eqnarray}
The multi-step function $w(x,y)$ appearing in the phase factors satisfies $w(x,y)=-w(y,x)=1$ for $x>y$, with $w(x,x)=0$.  
In terms of these operators the hamiltonian describing $N$ anyons of atomic mass $m$ confined in a length $L$ is
\begin{eqnarray}
H& =&\frac{\hbar^2}{2m} \int_0^L \rmd x \, \Psi^{\dagger}_x(x) \Psi_x(x)  + 
\frac12 \, g \int_0^L \rmd x \, \Psi^{\dagger}(x) \Psi^{\dagger}(x) \Psi(x) \Psi(x)  \nonumber\\
& &+\rmi \epsilon\int_0^L \rmd x \, [ \Psi^{\dagger}(x) \Psi^{\dagger}(x) \Psi(x) 
  \Psi_x(x) - \Psi_x^{\dagger}(x) \Psi^{\dagger}(x) \Psi(x) \Psi(x) ]
\label{Ham-new}
\end{eqnarray}
where we also impose periodic boundary conditions.

The tunable parameters in the model are (i) $g={\hbar^2c}/{m}$, the
dynamical interaction strength which drives the particles in elastic
scattering through zero range contact interaction, where $c$ is the coupling constant, 
(ii) the nonlinear
dispersion coupling constant $\epsilon$, which describes a delta shift in collisions, and (iii) 
the statistical parameter $\kappa$ which varies the statistics of the particles
from pure Bose statistics ($\kappa=0$) to pure Fermi statistics ($\kappa = \pi$).
The model differs from the interacting 1D Bose gas \cite{LL, McG} and 1D 
interacting anyons \cite{Kundu} due to the presence of the nonlinear
dispersion $\epsilon$, i.e.,  the s-wave scattering is disturbed by a delta shift in collisions.
Effectively, the 1D scattering length is increased or decreased in
individual pairwise collisions by this dispersion.
It follows that the model contains well known models as special cases in certain  limits.
For $\kappa=0$, hamiltonian (\ref{Ham-new}) reduces to that of the
generalized nonlinear Schr\"{o}dinger model \cite{Gutkin,Shnirman,Sen}, 
which has been studied by means of
scattering Bethe ansatz in the context of soliton physics.
For $\epsilon=0$, it describes the model of 1D interacting anyons.
%
For $\kappa=\epsilon=0$, it reduces to the model of 1D interacting bosons, while for
$\kappa=c=0$ it reduces to quantum derivative nonlinear Schrodinger model \cite{KB}.
As we shall see, the general model (\ref{Ham-new}) exhibits
more exotic dynamics than the two models corresponding to the special cases
$\kappa=0$ and $\epsilon=0$. 

For convenience, we hereafter set $\hbar =2m=1$.
We also use a dimensionless coupling constant $\gamma =c/n$ to characterize 
different physical regimes of the anyon gas, where $n=N/L$ is the linear density.

\section{Bethe Ansatz}

We first present the corresponding equation of motion 
$-\rmi \, \partial _t \Psi (x,t) =\left[H, \Psi (x,t) \right]$ via the
time-dependent quantum fields, i.e., the nonlinear Schr\"{o}dinger
equation is given by
\begin{equation}
\rmi  \Psi_t  = - \Psi_{xx} + g \Psi^{\dagger} \Psi \Psi +4 \rmi \epsilon \, \Psi^{\dagger} \Psi \Psi_x. \label{NSE}
\end{equation}
We note that this is an anyonic version of the generalized nonlinear
Schr\"{o}dinger equation. Restricting to its $N$-particle sector we can
reduce the eigenvalue problem of the hamiltonian (\ref{Ham-new}) to a
quantum many-body problem. 
To this end, we define a Fock vacuum state $\Psi(x)|0\rangle =0$. 
Thus we can prove that the number operator $N$ and
momentum operator $P$ are conserved, where
\begin{eqnarray}
N &=&\int_{0}^{L} \rmd x \,  \Psi^{\dagger}(x) \Psi(x)\\
P &=&\rmi  \int_{0}^L \rmd x \, \Psi_x^{\dagger} \Psi(x).
\end{eqnarray}

In order to properly assign the anyonic phase $w(x_i,x_j)$ \cite{Kundu,BGO,BGH,Korepin-1} 
in the domain $x_1<x_2 < \ldots <x_N$, we write the $N$-particle eigenstate as
\begin{equation}
\mid \! \Phi \rangle=\int_0^L \rmd x^N {\rme}^{-\rmi \frac12 {\kappa N}} 
\chi(x_1\ldots x_N)\Psi ^{\dagger}(x_1)\ldots \Psi^{\dagger}(x_N) \! \mid \! 0\rangle
\label{state}
\end{equation}
where the wavefunction amplitude is of the form  
\begin{eqnarray}
\chi(x_1 \ldots x_N)&=& {\rme}^{- \rmi  \frac12 \kappa \sum_{x_i<x_j}^Nw(x_i,x_j)}
\sum_PA(k_{P1}\cdots k_{PN}) {\rme}^{\rmi (k_{P1}x_1+\ldots +k_{PN}x_N)}.
\label{wave}
\end{eqnarray}
Here the sum extends over all $N!$ permutations $P$. 
The choice of the sign of the anyonic phase factor in equation \eref{wave}
can be arbitrary and may lead to different boundary conditions \cite{Korepin-1}. 
Here we prefer to a fix a boundary condition where we count  the anyonic phase factor in equation \eref{wave} 
through the order of the particles in the scattering states \cite{BGH}.  
By acting on the eigenstate \eref{state} with
hamiltonian \eref{Ham-new} the eigenvalue problem for hamiltonian \eref{Ham-new}, namely
\begin{eqnarray}
H\mid \! \Phi \rangle=\int_0^L \rmd x^N {\mathrm
  e}^{-\rmi \frac12{\kappa N}}
H_N\chi(x_1\ldots x_N)\Psi ^{\dagger}(x_1)\ldots \Psi^{\dagger}(x_N) \! \mid \! 0\rangle
\end{eqnarray}
can be reduced to solving the quantum mechanical problem
\begin{equation}
H_N\chi (x_1\ldots x_N)=E\chi (x_1\ldots x_N) \label{Q-E}
\end{equation}
where
\begin{eqnarray}
H_N&=&- \sum_{i = 1}^{N}\frac{\partial^2}{\partial x_i^2}+g \sum_{1\leq i<j\leq N}\delta(x_i-x_j)\nonumber\\
&& +\,\rmi  \, 2 \epsilon \sum_{1\leq i<j\leq N} \delta
(x_i-x_j)(\partial _{x_i}+\partial _{x_j}). \label{Ham-N}
\end{eqnarray}

This anyon model turns out to be Bethe ansatz solvable with independent values of $c$, $\epsilon$ and $\kappa$.  
Changing the coordinates to centre of mass coordinates 
$X=({x_j+x_k})/{2}$ and $Y=x_j-x_k$ leads to the eigenvalue equation (\ref{Q-E}) 
in the form
\begin{eqnarray}
& &\left\{ \left(-\frac{\partial^2}{\partial x_1^2}\cdots
-\frac{1}{2}\frac{\partial^2}{\partial
  X^2}-2\frac{\partial^2}{\partial Y^2}\cdots -\frac{\partial
  ^2}{\partial x_N^2}\right)\right.\nonumber \\
& &\left. \phantom{\frac{\partial^2}{\partial x_1^2}}          +2c \delta (Y)+ \rmi \, 4 \epsilon
\delta(Y)\partial _X-E \right\} \chi(\ldots x_i \ldots x_j \ldots )=0.
\end{eqnarray}
Integrating both sides of this  equation with respect to $Y$ from
$-e$ to $+e$ and taking the limit ${e \to 0}$ 
gives the discontinuity condition
\begin{eqnarray}
& &\left( \partial _{x_{j}}-\partial _{x_i}
\right)\chi(x_1,\ldots,
x_i,x_{j},\ldots,x_N)|_{x_j=x_i+e} \nonumber\\
&&-\left( \partial _{x_j}-\partial _{x_i}
\right)\chi(x_1,\ldots,
x_j,x_{i},\ldots,x_N)|_{x_j=x_i-e}\nonumber\\
& &
=[2c+\rmi \, 2\epsilon (\partial _{x_i}+\partial _{x_j})] \chi(x_1,\ldots, x_i,x_{j},\ldots,x_N)|_{x_i=x_j}
\label{jump}
\end{eqnarray}
on the derivative of the  wavefunction.
This equation gives a relation between the coefficients $A(k_{P1}\ldots k_{PN})$ 
of the form 
\begin{equation}
A(\ldots k_j,k_i
\ldots)=\frac{k_j-k_i+\rmi (c'-\epsilon'(k_j+k_i))}{k_j-k_i-\rmi (c'-\epsilon'(k_j+k_i))}
A(\ldots k_i,k_j \ldots)
\end{equation}
which is the two-body scattering relation.  
This relation gives rise to a functional scalar scattering matrix which factorizes three-body scattering 
processes into the product of three two-body scattering matrices \cite{Gutkin}.
In this sense the Yang-Baxter equation is trivially satisfied, guaranteeing 
no diffraction in the scattering process \cite{McG,Gutkin}.
Consequently the $N$-particle scattering amplitude can be factorized
into the products of two-particle amplitudes. 
Another way to establish the integrability of the model is to explicitly construct an 
infinite number of higher order conserved quantities, as shown for the 1D Bose gas in Ref.~\cite{Gutkin2}.

The effective interaction strengths  $c'$ and $\epsilon'$ appearing in the solution are given by 
\begin{equation}
c'=\frac{c}{\cos(\kappa/2)} \qquad {\rm and} \qquad \epsilon'=\frac{\epsilon}{\cos(\kappa/2)}.
\label{effective}
\end{equation}
In this model these effective coupling constants  implement the transmutation
between statistical and dynamical interactions.

To complete the solution in terms of the Bethe ansatz we impose the periodic boundary condition 
$
\chi(x_1=0, x_2\ldots , x_N)= \chi(x_2, \ldots , x_N, x_1=L)\label{PBC-2}
$
on the anyonic wavefunction in the fundamental region  $0 \le x_1<x_2 < \ldots <x_N \le L$. 
As in the solution of the interacting anyon model ($\epsilon = 0$), 
this leads to the appearance of various phase factors \cite{BGH,Korepin-1}.
The wave functions for other regions can be determined through application of  the anyonic symmetry.
The periodic boundary condition can also be imposed on other wavefunction
arguments $x_i$, in each case leading to the same anyonic phase factor in the Bethe
ansatz equations (BAE) \cite{BGH,Korepin-1}.
For this model, the energy is given by 
\begin{equation}
E=\sum_{i=1}^Nk_i^2
\end{equation}
where the quasi-momenta $k_j$ satisfy the BAE
\begin{equation}
{\mathrm e}^{\rmi k_jL}=-{\mathrm e}^{\rmi \kappa(N-1)} \prod^N_{\ell = 1}
\frac{k_j-k_\ell+\rmi \,(c'- \epsilon'(k_j+k_\ell))}{k_j-k_\ell-\rmi \, (c'-\epsilon'(k_j+k_\ell))}
\label{BA-new}
\end{equation}
for $j = 1, \ldots, N$. The total  momentum is 
\begin{equation}
p=\sum_{i=1}^Nk_i=N(N-1)\kappa/L.
\end{equation}
In minimizing the energy we consider $\kappa(N-1) = \nu$ (mod $2\pi$)
in the phase factor.

The Bethe ansatz solution provides in principle the full physics of the model. 
For the case $\kappa=0$, the BAE \eref{BA-new} are
consistent with the scattering states constructed for the generalized
nonlinear Schr\"{o}dinger model \cite{Gutkin,Shnirman}.  It is also clear
to see from the BAE \eref{BA-new} that the phase shift
\begin{equation}
\theta_{j,\ell}=\ln \left[ \frac{k_j-k_\ell+\rmi \,(c'-
\epsilon'(k_j+k_\ell))}{k_j-k_\ell-\rmi \,
(c'-\epsilon'(k_j+k_\ell))} \right]
\end{equation}
depends essentially on the anyon parameter $\kappa$, the dynamical interaction strength $c$ 
and the nonlinear dispersion of velocity $\epsilon$ through the effective interactions \eref{effective}.
Among the competing interactions, the 
nonlinearity  parameter $\epsilon$ is essential in determining the bound states. 
The $\delta$-function interaction strength $c$ plays the role of a driving force in collisions. 
However, the anyonic parameter $\kappa$ determines the statistical signature of the particles. 

\section{Preliminary analysis}

The asymptotic behaviour of the BAE (\ref{BA-new}) reveals that 
bound states exist for $c<0$ and $c<2\epsilon \nu/L$. 
In general, the form of the bound states appears to be extremely complicated. 
For example,  for $N=2$ and $L(c-p\epsilon)\gg 1$, we have the bound
state $\left\{\frac12 p \pm \rmi  \frac12 (c-p\epsilon) \right\}$. 
In this communication we focus on the repulsive regime, where $c>0$
and $|\epsilon| < c/(2p)$, for which the Bethe roots are real.

\begin{figure}
  \centering
  \vspace{10mm}
  \scalebox{0.60}{\includegraphics{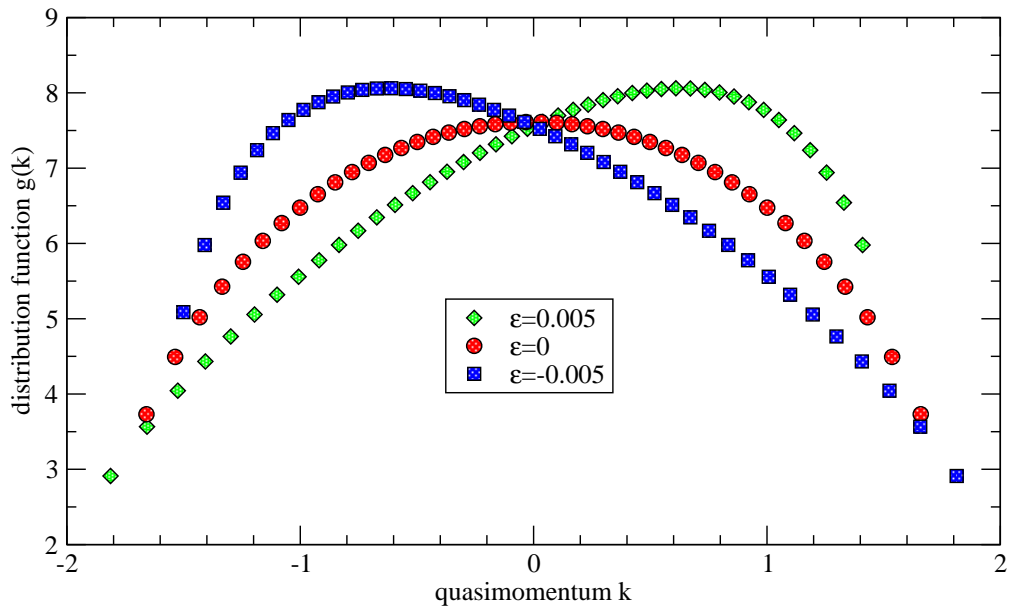}}
  \caption{Typical figure showing the quantum drift effect of the nonlinear dispersion $\epsilon$ on the 
quasimomentum distribution function. Here $\kappa=0$, $c = 0.04$, $L=2$ and $N=45$.  
Drifts to the left and right are symmetric with respect to the sign of $\epsilon$.}
\label{fig:distr-w0}
\end{figure}

In the weak coupling limit $\gamma \ll \cos(\kappa/2)$, the leading
term for the ground state energy, obtained from the BAE (\ref{BA-new}), is
\begin{equation}
\frac{E_0}{N} \approx \frac{(N-1)}{L}\left(c'-2\epsilon'\nu/L\right)+ \,\frac{\nu^2}{L^2}.
\end{equation} 
To this order, for $\kappa=0$ the ground state energy is independent of 
the  nonlinear dispersion $\epsilon$ and reduces to that of weakly interacting bosons.
However, the quasimomentum distribution 
\begin{equation}
g( (k_j+k_{j+1})/2 ) = \frac{1}{L(k_{j+1} - k_{j})}.
\end{equation} 
deviates from the usual semicircle law \cite{LL,Gaudin,BGM} 
at $\epsilon=0$ (see \fref{fig:distr-w0}). 
For $\epsilon >0$, the high density distribution drifts to the right hand side and the front becomes steeper as 
the nonlinear dispersion $\epsilon$  increases. 
For $\epsilon <0$, the high density part drifts to the left hand side. 
The nature of this kind of quantum drift in quasimomentum space
is due to the delta shift or equivalently the nonlinear dispersion of velocity, which
effectively increases or decreases the 1D scattering length in
individual pairwise collisions.

\begin{figure}
  \centering
  \vspace{10mm}
  \scalebox{0.60}{\includegraphics{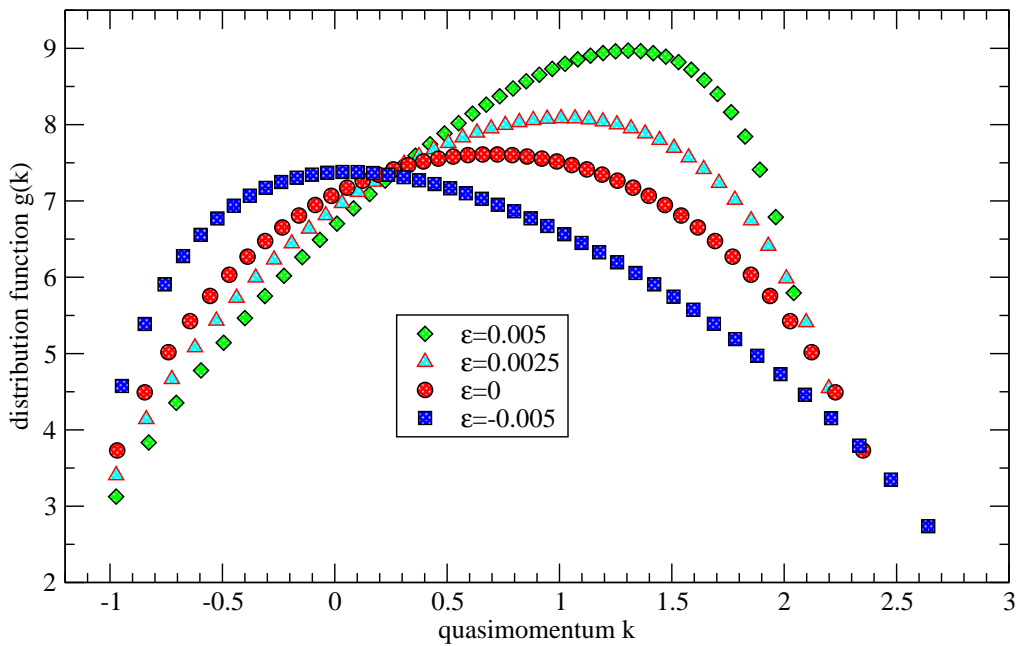}}
  \caption{Typical figure showing the quantum drift effect in the quasimomentum distribution function 
  with nonlinear dispersion $\epsilon$ and anyonic parameter $\kappa$. 
Here $\kappa={\pi}/{100}$, $c = 0.04$, $L=2$ and $N=45$.  
The drifts is no longer symmetric with respect to the sign of $\epsilon$.}
\label{fig:distr-wp}
\end{figure}

When the total momentum is non zero, i.e. $p\ne 0$, the left and right
drifts are no longer symmetric with respect to the sign of $\epsilon$.
Here the anyonic parameter $\kappa$ also introduces quantum drifts
in quasimomentum space. 
A particular example is illustrated in \fref{fig:distr-wp}. 
For $\epsilon>0$, the front of the distribution
become very steep and the width becomes narrower as $\epsilon$ increases.
The energy decreases as $\epsilon$ increases. 
However, for $\epsilon <0$, the distribution becomes flatter.
In this case, the energy increases as $|\epsilon|$ increases.

In the strong coupling limit $Lc \gg 1$, one can perform the 
strong coupling expansion with the BAE (\ref{BA-new}).
In this limit the ground state of the model becomes that of the Tonks-Girardeau gas.  
This is mainly because in the strong repulsive limit, the particles behave like 
fermions in thermalized states. 
As a result the effect induced by the nonlinear dispersion of velocity is suppressed. 
To obtain the explicit result for the ground state energy, let all quasimomenta $k_j$ shift to 
$\lambda _j=k_j-\nu/L$ and for  simplicity, choose $N$ to be odd. 
In this limit, the quasimomenta are given explicitly by
\begin{eqnarray}
\fl
\lambda _j&\approx &\left[\frac{2n_j\pi}{L}\left(1-\frac{4n_j\epsilon'\pi}{(Lc'-\epsilon'
  \nu)^2}\right) +\frac{2\epsilon'N(N^2-1)\pi^2}{3L(Lc'-\epsilon'\nu)^2}\right] 
  \left(1+\frac{2N}{(Lc'-\epsilon'  \nu)}  \right)^{-1}
\label{TG-n}
\end{eqnarray}
where $n_j=-(N-1)/2,\ldots, (N-1)/2$. 
This result indicates that $\sum_{j=1}^N\lambda_j=0$.
The corresponding ground state energy is given by
\begin{equation}
\frac{E_0}{N}\approx \frac{(N^2-1)\pi^2}{3L^2}\left(1+\frac{2N}{(Lc'-\epsilon'  \nu)}  \right)^{-2}
+\,\frac{\nu^2}{L^2}.
\end{equation}

It can be seen from the Bethe roots (\ref{TG-n}) that the flat fermion-like
distribution becomes inclined as $|\epsilon|$ becomes large. 
The distribution almost linearly increases or decreases depending on the
direction of the current and the parameters $\kappa$ and $\epsilon$. 
The system is strongly collective in the Tonks-Girardeau limit.

\section{Concluding remarks}

We have presented an exactly solved model of 1D anyons with general 
$\delta$-function and derivative $\delta$-function interaction.  
The Bethe ansatz solution has been obtained by means of the coordinate Bethe ansatz.
We have seen that the anyonic signature is inextricably
related to the nonlinear dispersion of velocity and the pairwise
dynamical interaction between identical particles.  
Competing interactions among the anyonic parameter $\kappa$ and the strengths of the
$\delta$-function interaction $c$ and the 
nonlinear dispersion of velocity $\epsilon$ result in more subtle bound states
and scattering states. 
Preliminary analysis of the Bethe Ansatz solution in the weak and strong coupling limits 
has revealed  drifts in the  quasimomentum distribution as a function of the 
nonlinear dispersion and the anyonic parameter.

The quasimomentum distribution may provide a
plausible way to observe anyonic behaviour in a general model with
$\delta$-function and derivative $\delta$-function interaction.
In particular, it suggests the possibility of observing anyonic behaviour 
via the generalized nonlinear Schr\"odinger equation, perhaps with regard
to the propagation of solitons in nonlinear media.
On the other hand, the observed quantum drifts in distribution may perhaps also be seen in
experiments with trapped ultracold atoms in and out of equilibrium \cite{T-G,Weiss}. 
The Bethe ansatz solution should also provide an accessible
way to explore the dynamics in time evolution of the generalized nonlinear Schr\"{o}dinger
equation (\ref{NSE}), just as the time-dependent dynamics of the 1D boson model has
recently been investigated \cite{Buljan}.
In addition, the investigation of fractional exclusion statistics \cite{Haldane,BGO,BG,Wu,Ha,NW,Wadati} 
in this generalized 1D anyon gas would provide insight into the influence on the statistical signature  
of both dynamical interaction and nonlinear dispersion.

\ack This work has been supported by the Australian Research Council.
The authors thank Chao-Heng Lee, Jing-Song He and Miki Wadati  for
some helpful discussions.  One of the authors (A. K.) thanks the
Department of Theoretical Physics at The Australian National
University for hospitality during his visit in Nov/Dec 2007.


\section*{References}
\begin{thebibliography}{10}

\bibitem{cond1} Camino F E, Zhou W  and  Goldman V J 2005  \PR B {\bf 72} 075342 
\nonum Kim E-A, Lawler Vishveshwara M S and Fradkin E 2005 \PRL  {\bf 95} 176402
\nonum Bonderson P, Kitaev A and Shtengel K 2006 \PRL {\bf 96}  016803

\bibitem{QFHE} Halperin B I 1984 \PRL {\bf 52} 1583
\nonum Wilczek F  ed. {\it Fractional Statistics and Anyon Superconductivity} 
(World Scientific, Singapore, 1990)

\bibitem{comp1} Das Sarma S, Freedman M and Nayak C 2005
  \PRL {\bf 94}  166802
\nonum  Bonesteel N E, Hormozi L and  Zikos G 2005 \PRL {\bf 95} 140503

\bibitem{comp2} Weeks C, Rosenberg G Seradjeh B and Franz M 2007 {\it Nature  Phys.} {\bf 3} 796

\bibitem{1D-anyon1}
Amico L, Osterloh A and Eckern U 1998  \PR B {\bf 58}  R1703
\nonum Osterloh A, Amico L and Eckern U 2000  \JPA {\bf 33} L487

\bibitem{1D-anyon2} Dukelsky J, Esebbag C and Schuck P 2001 \PRL {\bf 87} 066403

\bibitem{1D-anyon3} Liguori A and  Mintchev M 2000 \NP B {\bf 569} 577

\bibitem{Kundu} Kundu A 1999 \PRL {\bf 83} 1275

\bibitem{LL}Lieb E H and  Liniger W 1963  \PR {\bf 130} 1605

\bibitem{McG} McGuire J B 1964 \JMP {\bf 5} 622

\bibitem{BGH} Batchelor M T and Guan X-W and He J-S 2007 {\it J. Stat. Mech.} P03007
  
\bibitem{Korepin-1} Patu O I, Korepin V E and Averin D V 2007 
{\it J. Phys. A: Math. Theor.} {\bf 40} 14963
  
\bibitem{Zhu} Zhu R-G and Wang A-M, {\em Theoretical construction of 1D anyon models}, arXiv:0712.1264

\bibitem{Haldane} Haldane F D M 1991 \PRL {\bf 67} 937

\bibitem{BGO} Batchelor M T, Guan X-W and Oelkers N 2006 \PRL {\bf 96} 210402
  
\bibitem{BG} Batchelor M T and Guan X-W 2006 \PR B {\bf 74} 195121

\bibitem{Korepin-2} Patu O I, Korepin V E and Averin D V 2008
  {\em J. Phys. A: Math. Theor. } {\bf 41} 145006
  
\nonum Patu O I, Korepin V E and Averin D V 2008  {\em J. Phys. A: Math. Theor. } {\bf 41} 255205

\bibitem{Satachiara1} Santachiara R and Calabrese P 2008 {\it J. Stat. Mech.} P06005

\bibitem{Calabrese} Calabrese P and  Mintchev M 2007 \PR B {\bf 75} 233104

\bibitem{Satachiara2} Santachiara R, Stauffer R F and Cabra D 2007  {\it J. Stat. Mech.} L05003

\bibitem{delCampo} del Campo A {\em Fermionization and bosonization of expanding 1D anyonic fluids}, arXiv:0805.3786

\bibitem{HZC} Hao Y, Zhang Y and Chen S {\em Ground-state properties of one-dimensional anyon gases}, arXiv:0805.1988

\bibitem{AN} Averin D V and Nesteroff J A 2007 \PRL {\bf 99} 096801

\bibitem{Gutkin} Gutkin E 1987  {\it Ann. Phys.} {\bf 176}  22

\bibitem{Gutkin2} Gutkin E 1988  {\it Phys. Rep.} {\bf 167} 1

\bibitem{Shnirman} Shnirman A G, Malomed B A and Ben-Jacob E 1994
 \PR A  {\bf 50}  3453

\bibitem{Sen} Basu-Mallick B, Bhattacharyya T and Sen D 2005 \PL A {\bf 341} 371
 
\bibitem{KB} Kundu A and  Basu-Mallick B 1993 \JMP {\bf 34} 1252

\bibitem{Girardeau} Girardeau M D 2006  \PRL {\bf 97} 210401

\bibitem{Gaudin} Gaudin M {\em La fonction d'Onde de Bethe} (Masson, Paris, 1983) 

\bibitem{BGM} Batchelor M T, Guan X-W and McGuire J B 2004 \JPA {\bf 37} L497

\bibitem{T-G} Paredes B, Widera A, Murg V, Mandel O, Folling S, Cirac I, Shlyapnikov GV, Hansch TW, Bloch I  
2004  {\it Nature} {\bf 429} 277

\bibitem{Weiss} Kinoshita T,  Wenger  T and Weiss D S 2006 {\it Science} {\bf 305} 1125
  \nonum  Kinoshita T, Wenger T and  Weiss D S 2006 {\it Nature} {\bf 440} 900

\bibitem{Buljan} Buljan H, Pezer R and  Gasenzer T 2008 \PRL {\bf 100} 080406

\bibitem{Wu} Wu Y-S  1994 \PRL {\bf 73} 922

\nonum Bernard D and Wu Y-S in {\em New Developments of Integrable Systems and Long-Ranged 
Interaction Models} Ge M-L and Wu Y-S eds (World Scientific, Singapore, 1995) p 10

\bibitem{Ha} Ha Z N C 1994 \PRL {\bf 73} 1574

\bibitem{NW} Nayak C and Wilczek F 1994 \PRL {\bf 73} 2740 

\bibitem{Wadati} Wadati M 1985 {\it J. Phys. Soc. Japan} {\bf 54} 3727
\nonum Wadati M 1995 {\it J. Phys. Soc. Japan} {\bf 64} 1552

\end{thebibliography}
\end{document}